\documentclass[journal]{IEEEtran}

\usepackage[cmex10]{amsmath}
\usepackage{amsfonts,amssymb}
\usepackage{bm}
\usepackage{cite}
\usepackage{booktabs}
\usepackage{array}
\usepackage{algorithm}
\usepackage{algpseudocode}
\usepackage{xcolor}
\usepackage{tikz}
\usetikzlibrary{arrows.meta,positioning,fit,calc}
\usepackage{hyperref}
\hypersetup{hidelinks}

\graphicspath{{Graphics/}}

\newcommand{\CN}{\mathcal{CN}}

\newcommand{\Cplx}{\mathbb{C}}
\newcommand{\Ex}{\mathbb{E}}

\newcommand{\Herm}{{\mathsf{H}}}

\newcommand{\sigmoid}{\operatorname{sigmoid}}

\graphicspath{{Graphics/}}
\definecolor{ColorHighlight}{rgb}{1,0,0}

\begin{document}

\title{Learning-to-Transition for Large-scale and High-Order MIMO Detection}

\author{Yubo Zhang, Yiyao Liu, and Xiaodong Wang}
\maketitle

\begin{abstract}
High-order multiple-input multiple-output (MIMO) detection requires efficient search over a large discrete symbol space while producing reliable soft information for channel decoding. This paper develops a learning-to-transition (L2T) framework that formulates MIMO detection as a stochastic sequence of complete-vector transitions. At each transition, a channel-coupled Transformer updates both the instance embedding and the sampling policy, while a blockwise autoregressive factorization captures inter-stream dependence with moderate sequential complexity. For hard-output detection, a transition network is applied recursively and trained through a residual-to-BER curriculum, which first learns the MIMO search geometry from the exact residual metric and then aligns the policy with transmitted-bit accuracy. For soft-output reception, the well-trained hard policy is cloned at the parameter level into every layer of an untied soft-input soft-output iterative detection and decoding (IDD) receiver. This tied-to-untied transfer preserves the learned zero-prior search dynamics while enabling layer- and round-specific specialization under decoder feedback. Within each IDD round, decoder priors tilt candidate generation according to Bayes' rule, and likelihood-weighted terminal hypotheses produce posterior and extrinsic log-likelihood ratios for LDPC decoding. A multi-stage training strategy further stabilizes the hard-to-soft transfer by progressively exposing the receiver to synthetic and in-loop decoder-generated priors.
\end{abstract}


\section{Introduction}\label{sec:introduction}

Multiple-input multiple-output (MIMO) transmission improves spectral efficiency
by multiplexing several data streams, but joint detection becomes increasingly difficult as
the system loading and modulation order grows. With perfect channel state
information (CSI), maximum-likelihood (ML) detection must search over $Q^{N_t}$ possible
transmit vectors for $N_t$ streams and a $Q$-ary constellation. This exponential
complexity is prohibitive for large or square high-order MIMO systems, whereas
low-complexity linear detectors may suffer substantial performance loss under strong
inter-stream interference or ill-conditioned channels \cite{yang2015fifty}.

Approximate message passing (AMP), orthogonal AMP, expectation propagation (EP), and model-driven unfolding methods provide useful performance-complexity tradeoffs. Learned receivers, including OAMP-Net2, RE-MIMO, and the Soft Graph Transformer (SGT), further improve robustness or better capture dense inter-stream interactions~\cite{he2020model,pratik2021remimo,hong2025sgt}. Nevertheless, most existing neural detectors either inherit a prescribed inference recursion or directly map a received instance to symbol estimates or marginals. As a result, they do not explicitly learn how to traverse the discrete space of complete high-order MIMO transmit vectors. Related combinatorial and Ising formulations expose the same discrete structure, but typically rely on specialized optimization procedures rather than an amortized transition policy~\cite{norimoto2023hubo,roychowdhury2023oim}.

Reinforcement learning for combinatorial optimization offers a complementary perspective. Constructive policies generate a solution one component at a time, whereas learned improvement heuristics usually apply a prescribed local modification~\cite{bello2017nco,wu2021improvement}. In contrast, we learn the stochastic transition rule itself. Across search steps, the same Transformer is applied recursively to map the current complete MIMO vector to a distribution over the next complete vector; within each step, the next vector is generated in a blockwise autoregressive manner across transmit streams. The sampled candidates are then evaluated by the exact channel metric. Therefore, the term ``complete-vector transition'' refers to the search state before and after each transition, while the decisions within a transition can still follow a structured autoregressive factorization.

A second difficulty is objective alignment. The residual energy provides a model-based cost for every sampled vector, but minimizing the residual alone does not necessarily minimize the transmitted-bit BER. We therefore adopt a residual-to-BER curriculum for hard-detector training. The policy is first trained with a residual-driven policy-gradient objective, which teaches the coupled MIMO search geometry, and is then smoothly shifted to a transmitted-bit binary cross-entropy (BCE) objective, which directly aligns the learned policy with bit-error-rate performance~\cite{wiesmayr2023ber}.

Soft-output reception further requires reliable uncertainty quantification, not merely a strong hard estimate. Iterative detection and decoding (IDD) needs posterior and extrinsic soft information, together with competing hypotheses for both values of each coded bit. Classical SISO and variational detectors make this posterior/extrinsic distinction explicit~\cite{studer2008siso,lin2008variational}. We follow this principle by transferring the hard policy to a prior-conditioned soft detector. Specifically, the well-trained tied hard policy is cloned at the parameter level into every transition layer of every IDD round, and the resulting copies are then optimized independently. This tied-to-untied transfer preserves the learned zero-prior search dynamics at initialization, while allowing each soft layer to specialize to its position in the search trajectory and to the decoder-prior distribution of its IDD round. During soft-output inference, decoder priors tilt candidate sampling according to Bayes' rule; a likelihood-weighted terminal mixture produces detector posterior LLRs; and explicit prior subtraction yields the extrinsic LLRs passed to the decoder.

The main contributions are summarized as follows:
\begin{itemize}
  \item We formulate high-order MIMO detection as a complete-vector transition
  process and develop a channel-coupled Transformer policy for this process.
  Instead of constructing a transmit vector component by component or applying a
  prescribed local update, the detector learns to map the current complete vector
  to a distribution over the next complete vector. The Transformer jointly updates
  the instance embedding and the sampling policy, while a blockwise autoregressive
  factorization captures inter-stream dependence within each transition.

  \item We train the hard-output detector through trajectory-level reinforcement
  learning with a residual-to-BER curriculum. The residual-driven policy-gradient
  objective first exploits the exact channel metric to learn the coupled MIMO
  search geometry, and the later transmitted-bit BCE objective directly aligns the
  policy with bit-error-rate performance.

  \item We extend the hard detector to a soft-input soft-output IDD receiver via
  tied-to-untied hard-to-soft transfer. The well-trained tied hard policy is cloned
  at the parameter level into every independently trainable soft-detector layer,
  providing a common zero-prior search initialization while allowing specialization
  across transition depth and IDD rounds. The soft receiver further incorporates
  Bayesian prior-tilted sampling, likelihood-weighted posterior LLR computation,
  explicit extrinsic LLR generation, and staged training for drifting decoder
  priors.
\end{itemize}

\section{Problem Statement}\label{sec:problem}


\subsection{MIMO Detection}\label{mimo_detect_bg}

\begin{figure}
    \centering
    \includegraphics[width=0.85\linewidth]{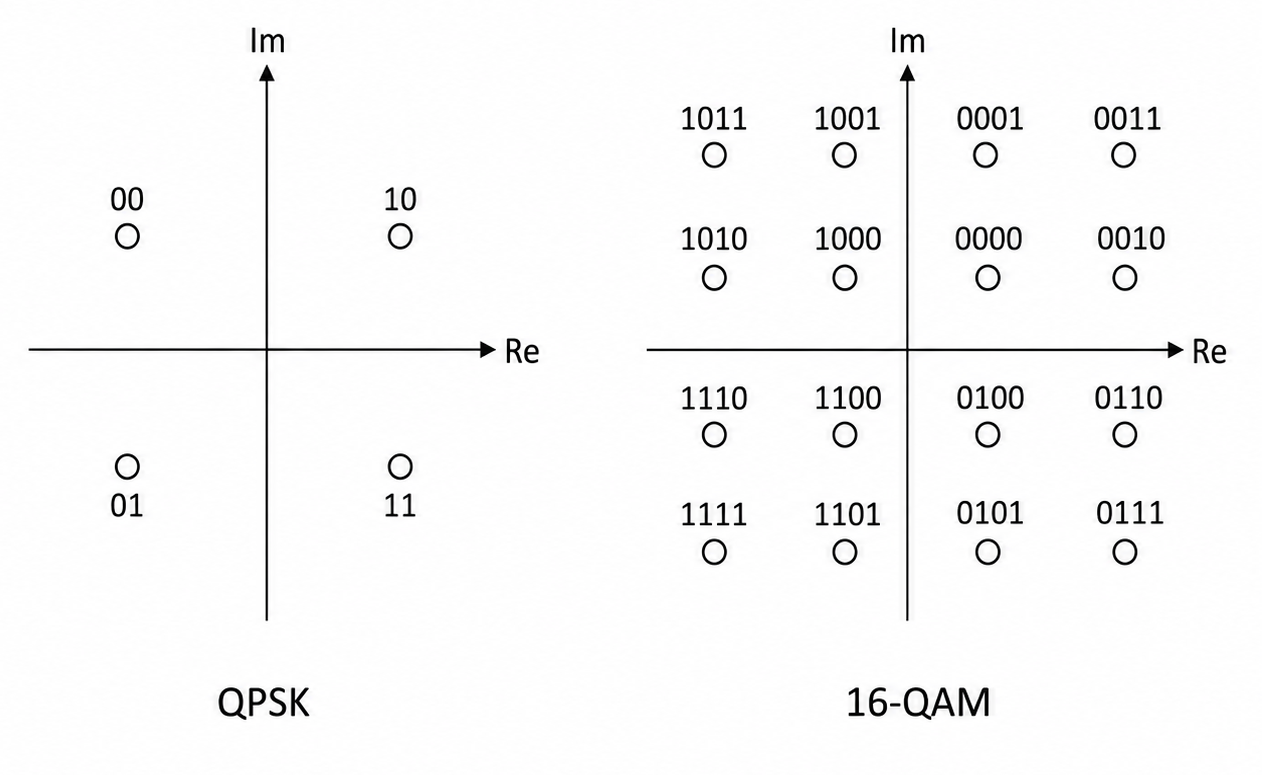}
    \caption{Constellation for Gray-coded data symbols.}
    \label{qam_constellation}
\end{figure}

Consider a point-to-point spatial-multiplexing MIMO channel with $N_t$
transmit antennas and $N_r$ receive antennas. For each channel use,
\begin{equation}\label{system}
\bm y=\bm H\bm x+\bm n,
\quad
\bm n\sim\CN(\bm 0,\sigma^2\bm I_{N_r}),
\end{equation}
where $\bm H\in\Cplx^{N_r\times N_t}$ is the channel matrix, following the distribution $\bm{H} \sim p_H$, and $\sigma^2$ is the noise variance at each receive antenna. Let $\mathcal A_Q$ denote a unit-energy, Gray-coded quadrature-amplitude modulation (QAM) constellation with order $Q$. Each QAM symbol contains $B=\log_2Q$ coded bits, with $B/2$ bits on each pulse-amplitude modulation (PAM) axis. The corresponding constellation mapping is depicted in Fig.~\ref{qam_constellation}, demonstrating how QAM scheme exploits the two-dimensional space to separate modulated symbols.  

Let $\mu_Q:\{0,1\}^{B}\rightarrow\mathcal A_Q$ be the prescribed Gray
mapping. We collect the bits assigned to all transmit antennas and define the modulated symbol as follows:
\begin{equation}\label{mapping}
\bm x=\mathcal{M}_Q(\bm b)
=
\big[
\mu_Q(\bm b_1),\ldots,\mu_Q(\bm b_{N_t})
\big]^{\mathsf T},
\end{equation}
where
$\bm b=[\bm b_1^{\mathsf T},\ldots,\bm b_{N_t}^{\mathsf T}]^{\mathsf T}$
belongs to $\mathcal B=\{0,1\}^{N_tB}$. Traditional maximum-likelihood (ML) detection then adopts the following exhaustive search:
\begin{equation}\label{hard_ml_rule}
\hat{\bm b}_{\mathrm{ML}}(\bm{y},\bm{H}) = \arg\min_{\bm b\in\mathcal B} |\bm y-\bm H\mathcal{M}_Q(\bm b)\|_2^2.
\end{equation}
Note that the entire space now contains $2^{N_tB}$ feasible solutions, hence the ML detection in \eqref{hard_ml_rule} becomes computationally prohibitive under the large system size (i.e., $N_t$ increases) and high-order QAM (i.e., $B$ increases). Moreover, it is inefficient to repeatedly solve similar detection problems as the channel $\bm{H}$ frequently varies.

To tackle these issues, we aim to train a single neural network that solves the symbol detection problems for channel instances across the entire distribution $\bm{H} \sim p_H$. This follows the essence of learning-to-optimize (L2O)/ amortized optimization, which shifts the computational burden from online iterative optimization to offline training. Instead of solving each problem instance from scratch, L2O learns an optimizer from the data, exploiting recurring task structures for fast inference. Specifically, denote the residual function of the symbol detection as follows:
\begin{align}\label{residual_cost}
f(\bm b;\bm{y},\bm{H}) = \|\bm y-\bm H\mathcal{M}_Q(\bm b)\|_2^2.
\end{align}
Then the L2O solver aims to minimize this metric in the population level: let $\hat{\bm b}_{\bm\theta}(\bm{y},\bm{H})$ denotes its final output, where $\bm\theta$ denotes the learnable parameters of the neural solver, the amortized learning problem becomes
\begin{align} \label{amortizedproblem}
\min_{\bm\theta} \ \Ex_{\bm{H} \sim p_H} \left[ f \big(\hat{\bm b}_{\bm\theta}(\bm{y},\bm{H});\bm{y},\bm{H}\big) \right].
\end{align}
This formulation remains challenging, since symbol detection is inherently an NP-hard combinatorial optimization problem, which lacks a high-quality general-purpose solver.

\subsection{Learning to Transition}\label{recur_rl_solver}


Based on the objective in \eqref{amortizedproblem}, we seek to train a neural network that learns an effective transition policy over the entire solution space $\mathcal{B}$. Similar to the prior graph optimization work, learning an expressive channel embedding is crucial to the detection performance in this work. Note that the residual function in \eqref{residual_cost} can be rewritten as
\begin{equation}\label{quadraticmetric}
\|\bm y-\bm H\bm x\|_2^2
=
\bm x^{\Herm}\bm R\bm x
-2\operatorname{Re}\{\bm x^{\Herm}\bm z\}
+\|\bm y\|_2^2,
\end{equation}
where $\bm{x} \triangleq \mathcal{M}_Q(\bm b) \in \mathbb{C}^{N_t}$ represents the modulated symbol, $\bm R \triangleq \bm H^{\Herm}\bm H \in \mathbb{R}^{N_t \times N_t}$ represents the Gram/correlation matrix of channel $\bm{H}$, and $\bm z \triangleq \bm H^{\Herm}\bm y \in \mathbb{C}^{N_t}$ represents the corresponding matched filter output. Inspired by~\cite{khalil2017learning}, the effective channel features can be learned by drawing \textbf{an analogy with graph-embedding generation}. Specifically, the matched-filter entries, i.e., $\bm{z}$, and all diagonal entries of Gram matrix, i.e., $\text{diag}(\bm{R})$, form the node-wise inputs (each node relates to one transmit antenna); while all off-diagonal entries of Gram matrix, i.e., $\text{offdiag}(\bm{R})$, form the edge-wise inputs (which reflects the complex inter-stream couplings). Then a graph neural network (GNN) encoder $G_{\bm\varphi}$ maps these sufficient input statistics to the initial channel embedding, i.e., 
\begin{align} \label{initialembeddingmatrix}
\bm Z^{(0)} = G_{\bm\varphi} \big(\bm{z}, \text{diag}(\bm{R}), \text{offdiag}(\bm{R})\big) \in \mathbb{R}^{D \times N_t},
\end{align}
where $D$ is the feature dimension of each node (transmit side).


To reduce the difficulty of learning the transition policy, inspired by the recursive optimization process in~\cite{johnston2023curriculum,zhang2025sallom}, we consider a \textbf{Learning-to-transition} (L2T) scheme consisting of $T$ updates, therefore amortizing the searching task across all steps. Importantly, a static embedding cannot reflect which parts of the search space have already been explored by the evolving candidate. We therefore keep updating the channel embedding across the transition steps, inspired by the works in \cite{khalil2017s2v,schulman2015scg}. Assume that the transition process at each step is performed by a neural network with learnable parameters $\bm\theta_{0}$. Hence the  
forward process at each transition step $t$ is given by
\begin{align}\label{jointtransitionoutput}
\left[ \bm p^{(t)}, \bm Z^{(t)} \right] = \Pi_{\bm\theta_{0}} \left( \bm b^{(t-1)}, \bm Z^{(t-1)} \right),
\end{align}
for $t=1,\dots,T$. The first output undergoes a bit-wise sampling process to obtain the next solution, i.e., $\bm b^{(t)}\sim\bm p^{(t)}$, while the second output is the updated feature embedding at transition step $t$.

\section{Hard-Decision Receiver}\label{hard_detector_sec}

We now design a non-iterative receiver where the transmitted bits are directly obtained via hard decisions. Equivalently, a learning-based solver is trained to address the NP-hard combinatorial optimization problem in a population level. To achieve this, we first propose the policy network design and then provide the policy gradient-based training scheme.

\subsection{Transformer-based Policy Network}

The transition policy $\Pi_{\bm\theta_0}$ in \eqref{jointtransitionoutput} can be parameterized by a Transformer network \cite{vaswani2017attention}. 
In particular, the previous sampled solution is reshaped as $\bm{b}^{(t-1)} = [\bm{b}^{(t-1)}_1,\dots,\bm{b}^{(t-1)}_{N_t}] \in \mathbb{R}^{B \times N_t}$, and then concatenated with the corresponding channel embedding $\bm{Z}^{(t-1)} \in \mathbb{R}^{D \times N_t}$ in a top-down manner, which yields the input token sequence to the next layer:
\begin{align}
\bm C^{(t)}=[\bm b^{(t-1)}; \bm Z^{(t-1)}] \in \mathbb{R}^{(B+D) \times N_t}.
\label{tokenmatrix}
\end{align}


Let $\bar{\bm C}^{(t)}=\text{LayerNorm}(\bm C^{(t)})$. For each attention head $h$, $h=1,\dots,H$, the query, key, and value projections are
\begin{align}
\bm Q^{(t,h)}
&=(\bar{\bm C}^{(t)})^{\top} \bm W_Q^{(h)} \in\mathbb{R}^{N_t \times D_k/H},
\notag\\
\bm K^{(t,h)}
&=(\bar{\bm C}^{(t)})^{\top} \bm W_K^{(h)} \in\mathbb{R}^{N_t \times D_k/H},
\notag\\
\bm V^{(t,h)}
&=(\bar{\bm C}^{(t)})^{\top} \bm W_V^{(h)} \in\mathbb{R}^{N_t \times D_v/H}.
\label{qkvprojection}
\end{align}
The corresponding attention output is
\begin{align} \label{attention_matrix}
\bm{A}^{(t,h)}
=\operatorname{softmax}\!\left(
\frac{\bm{Q}^{(t,h)}\bm{K}^{(t,h)\top}}{\sqrt{D_k/H}}
\right)\bm{V}^{(t,h)}.
\end{align}
The $H$ attention outputs are concatenated and projected to
\begin{align} \label{h_attn_head_output}
\bm{Y}^{(t)}
=\bm{W}_{o}[\bm{A}^{(t,1)},\ldots,\bm{A}^{(t,H)}]^{\top}
=[\bm{Y}_{p}^{(t)};\bm{Y}_{f}^{(t)}],
\end{align}
where $\bm{W}_{o} \in \mathbb{R}^{(B+D) \times D_v}$, $\bm{Y}_{p}^{(t)}\in\mathbb{R}^{B \times N_t}$ represents the solution logits and $\bm{Y}_{f}^{(t)}\in\mathbb{R}^{D \times N_t}$ contains the channel embedding residuals. Then the solution at the next transition step is sampled as follows:
\begin{align} \label{full_prob_output}
&\bm{P}^{(t)}=\text{sigmoid}(\bm{Y}_{p}^{(t)}) = [\bm{p}^{(t)}_1,\dots,\bm{p}^{(t)}_{N_t}],\notag \\
&\bm{b}_i^{(t)} \sim \operatorname{Bernoulli}(\bm{p}_i^{(t)}), \ i=1,\dots,N_t.
\end{align}
In parallel, the channel embedding is updated through residual connections and an MLP:
\begin{align} \label{rm_feature_update}
\tilde{\bm{Z}}^{(t)}
&=\bm{Z}^{(t-1)}+\bm{Y}_{f}^{(t)},\notag\\
\bm{Z}^{(t)}
&=\tilde{\bm{Z}}^{(t)}
+\operatorname{MLP}\!\left(
\operatorname{LayerNorm}(\tilde{\bm{Z}}^{(t)})
\right).
\end{align}
(\ref{full_prob_output}) and (\ref{rm_feature_update}) are the two output branches of the same Transformer layer. Unrolling this layer over $T$ transitions recursively refines both the discrete solution and the continuous embedding.

\subsection{Blockwise Autoregressive Sampling} \label{block_ar_sampling}

In principle, the mean-field sampling for each solution (i.e., every bit is sampled in parallel in the process $\bm{b}^{(t)} \sim \bm{p}^{(t)}$) is fast but ignores conditional correlations among $N$ bits. At the other extreme, bit-by-bit autoregressive sampling captures every conditional dependence but requires a large mount of sequential forward processes. We therefore introduce blockwise autoregressive sampling~\cite{stern2018blockwise} as an efficient tradeoff between these two extreme cases, to generate the complete solution $\bm{b}^{(t)}$ at each transition step. In this work, each block represents a $B$-bit QAM symbol from each transmit antenna. The dependence is retained across those $N_t$ symbols because the dense off-diagonal entries in the Gram matrix $\bm{R}$ couple the symbol detections at different antennas. In contrast, the bits within each QAM symbol are generated
in parallel. This mean-field factorization avoids a bit-by-bit chain of length $N_tB$ while preserving the dominant inter-symbol dependence.

The block ordering $n=1,2,\dots,N_t$ may be randomized during training to avoid imposing a persistent ordering bias. Define $\bm{b}_{\text{mask}}$ as an all-zero vector with length $B$, which can be regarded as each undetermined data symbol during a sequential sampling process. Then we define the following partial solution before sampling each data symbol $\bm{b}^{(t)}_{n}$ at transition step $t$:
\begin{align} \label{ar_partial_solution}
\bm{b}^{(t)}_{<n} \triangleq [\bm{b}^{(t)}_{1},\dots,\bm{b}^{(t)}_{n-1},\underbrace{\bm{b}_{\text{mask}},\dots,\bm{b}_{\text{mask}}}_{N_t-n+1 \ \text{copies}}]
\end{align}
Conditioned on $\bm{b}^{(t)}_{<n}$, the complete prior solution $\bm{b}^{(t-1)}$, and the current channel embedding $\bm{Z}^{(t)}_{n-1}$, the policy samples all variables in the next block in parallel:
\begin{align} \label{block_rollout_sample}
&[\bm{p}^{(t)}_n,\bm{Z}^{(t)}_n]
=\Pi_{\bm\theta_0}\!\left(
\cdot\mid \bm{b}^{(t)}_{<n},
\bm{b}^{(t-1)},\bm{Z}^{(t)}_{n-1}
\right),\notag\\
&\bm{b}_n^{(t)} \ \overset{\text{parallel}}{\sim} \ \operatorname{Bernoulli}(\bm{p}^{(t)}_n),
\end{align}
with $\bm{Z}^{(t)}_0=\bm{Z}^{(t-1)}$ and $\bm{Z}^{(t)}=\bm{Z}^{(t)}_{N_t}$. After $N_t$ blockwise samplings, we obtain the complete solution $\bm{b}^{(t)}$. Note that the special case $M=1$ reduces to a mean-field sampler, whereas $M=N$ reduces to a bit-by-bit autoregressive sampler.

\subsection{Policy-Gradient Training Scheme}

Implementing the above $T$-step stochastic transition process yields a complete L2T solution trajectory $\tau \triangleq (\bm b^{(0)},\bm b^{(1)},\ldots,\bm b^{(T)})$. Its sampling probability can be computed as
\begin{equation}\label{trajprob}
\begin{aligned}
p_{\bm\theta_{0}}(\tau\mid \bm{y}, \bm{H})
&=
p(\bm b^{(0)})
\prod_{t=1}^{(t)}
\Pi_{\bm\theta_{0}}
\left(
\bm b^{(t)}
\mid
\bm b^{(t-1)},\bm Z^{(t-1)} ; \bm{y}, \bm{H}
\right).
\end{aligned}
\end{equation}
where $p(\bm b^{(0)})$ does not depend on $\bm\theta_{0}$. Denote the step-wise objective function as $g(\bm b^{(t)};\bm{y}, \bm{H})$. Similar to the GCO paper, the following trajectory-wise objective function with entropy regularization is proposed:
\begin{align} \label{sampletrajloss}
g(\tau;\bm{y}, \bm{H})
&=
\sum_{t=1}^{T} g(\bm b^{(t)};\bm{y}, \bm{H}), \notag \\
J_{\lambda}(\bm\theta_{0};\tau,\bm{y}, \bm{H})
&=
g(\tau;\bm{y}, \bm{H})
+\lambda\log p_{\bm\theta_{0}}(\tau\mid\bm{y}, \bm{H}).
\end{align}
Since $\Ex_\tau[\log p_{\bm\theta_{0}}(\tau \mid \bm{y}, \bm{H})]$ is the negative entropy term, the coefficient $\lambda>0$ encourages explorations at the early stage. The corresponding distribution-level learning objective is then given as follows:
\begin{equation}\label{expectedobjective}
\mathcal J_{\lambda}(\bm\theta_{0})
=
\Ex_{\bm{H} \sim p_H}
\Ex_{\tau\sim p_{\bm\theta_{0}}(\cdot\mid\bm{y},\bm{H})}
\left[
J_{\lambda}(\bm\theta_{0};\tau,\bm{y},\bm{H})
\right],
\end{equation}
and the ultimate learning problem becomes 
\begin{equation} \label{opt_formu}
\min\limits_{\bm\theta_{0}} ~ \mathcal J_{\lambda}(\bm\theta_{0}),
\end{equation}
which is solved using a gradient-based method. Note that \eqref{expectedobjective} contains two nested expectations. The outer expectation $\Ex_{\bm{H} \sim p_H}$ arises because we aim to learn a population-based solver whose policy generalizes across the channel distribution; The inner expectation $\Ex_{\tau\sim p_{\bm\theta_{0}}(\cdot\mid\bm{y},\bm{H})}$ accounts for the stochasticity of the learned transition policy because each trajectory is sampled from the probability distributions output by the network. For each detection instance $(\bm{y},\bm{H})$, this expectation is approximated by averaging over multiple sampled trajectories $\{\tau_1,\dots,\tau_K\}$. Although the ultimate interest is $f(\bm{b}^{(T)};\bm{y},\bm{H})$ at the final step, aggregating the objective values along the trajectory encourages progressive improvement throughout the search process. 
At inference, for a new instance $(\bm{y},\bm{H})$, we expect the recursive application of the trained policy network to produce a near-optimal result $\bm{b}^{(T)}$. Since each update only requires a feedforward mapping and a sampling operation, the resulting solver can be executed repeatedly with low latency, making real-time deployment feasible.

We next show how the step-wise objective function $g(\bm b^{(t)};\bm{y}, \bm{H})$ is constructed. Note that the ultimate goal is to reduce the detection bit-error-rate (BER), and the true transmitted bits are assumed to be known during training. Hence the BER-oriented binary cross-entropy (BCE) can be regarded as an effective step-wise loss function: specifically, let $N_b=N_tB$ denote the entire codeword length; if the ground truth is $b_j^*$ for every $j^{\text{th}}$ bit, while the network produces probabilities $p^{(t)}_j \triangleq P^{(r)}_{\bm\theta_{0}}(b_j=1)$ at transition step $t$, $j=1,\dots,N_b$, the BER-oriented BCE loss for this codeword is  
\begin{align} \label{bit_bce_def}
\ell^{\mathrm{BCE}}_t= -\frac{1}{N_b} \sum_{j=1}^{N_b} \left[ b_j^* \log p^{(t)}_j +(1-b_j^*)\log(1-p^{(t)}_j) \right].
\end{align}
In practice, the residual loss function in \eqref{residual_cost} is also utilized as a smooth surrogate in the early-stage training. Hence the following hybrid step-wise loss function is proposed
\begin{align} \label{hybrid_step_loss_def}
g(\bm b^{(t)};\bm{y}, \bm{H}) = \rho_s \cdot f(\bm b^{(t)};\bm{y}, \bm{H}) + (1-\rho_s) \cdot \ell^{\mathrm{BCE}}_t,
\end{align}
where the coefficient $\rho_s$ gradually decreases from 1 to 0, explicitly reflecting a schedule shifts from the smooth residual loss to the BER-oriented BCE loss. As a result, substituting \eqref{hybrid_step_loss_def} into \eqref{sampletrajloss} yields the trajectory-wise loss function. 

According to \eqref{expectedobjective}, when computing the population-level loss, for each instance $(\bm{y},\bm{H})$, $K$ solution trajectories are sampled, which accounts for the expectation term $\Ex_{\tau\sim p_{\bm\theta_{0}}(\cdot\mid\bm{y},\bm{H})}$. The initializations of those $K$ trajectories (denoted as $\bm{b}_1^{(0)},\dots,\bm{b}_K^{(0)}$) are all randomly-generated valid solutions. Finally, the policy network $\Pi_{\bm\theta_0}$ is trained according to \eqref{opt_formu}.

During inference, for each $(\bm{y},\bm{H})$, we recursively employ the trained policy network $\Pi_{\bm\theta_0^*}$ for $T$ transitions based on the random initializations $(\bm{b}_1^{(0)},\dots,\bm{b}_K^{(0)})$, and obtain the final solution set $(\bm{b}_1^{(T)},\dots,\bm{b}_K^{(T)})$. The hard decision result is the final solution $\bm{b}_{k^*}^{(T)}$ with the smallest residual loss, i.e.,
\begin{equation}\label{hardout}
\bm{b}_{k^*}^{(T)}: k^*=\arg \min_{k=1,\dots,K} f(\bm b_k^{(T)};\bm{y},\bm{H}).
\end{equation}

\section{Soft-Input Soft-Output Receiver}\label{sec:soft}

\subsection{Iterative Detection and Decoding}

Iterative detection and decoding (IDD) exchanges soft information between the
MIMO detector and a soft-output low-density parity-check (LDPC) decoder.
An interleaver distributes one LDPC codeword over a batch of MIMO observations;
after detection, their LLRs are simply reassembled in codeword order before
decoding. We use the convention
$L(b)=\log\left(\frac{P(b=0)}{P(b=1)}\right)$, so $P(b=1)=\sigmoid(-L(b))$.
Let $\Gamma$ denote the code interleaver. At the $m^{\text{th}}$ IDD round, the complete turbo exchange uses the LDPC belief-propagation (BP) mapping $\mathcal D_{\mathrm{BP}}$, given as follows:
\begin{align}
\bm L^{\mathrm E,(m)}
&=
\bm L^{\mathrm P,(m)}-\bm L^{\mathrm A,(m)},
\label{detectorround}\\
\bm L_{\mathrm D}^{\mathrm{in},(m)}
&=
\Gamma^{-1}(\bm L^{\mathrm E,(m)}),
\
\bm L_{\mathrm D}^{\mathrm P,(m)}
=
\mathcal D_{\mathrm{BP}}(\bm L_{\mathrm D}^{\mathrm{in},(m)}),
\label{decoderround}\\
\bm L^{\mathrm A,(m+1)}
&=
\Gamma
\left(
\bm L_{\mathrm D}^{\mathrm P,(m)}
-\bm L_{\mathrm D}^{\mathrm{in},(m)}
\right),
\label{priorupdate}
\end{align}
where the subscripts $(A,P,E)$ in the above equations represent the a-priori, posterior, and extrinsic LLR. The exchange is initialized with $\bm L^{\mathrm A,(0)}=\bm0$.
The two subtractions are essential: each module passes only information newly
created from its own observation or parity constraints, rather than feeding its
input prior back as if it were new evidence.

\subsection{LLR Computations} \label{llr_compute}

For a bit $b$, Bayes' rule gives
\begin{equation}\label{bayesbit}
P(b\mid\mathcal F,L^{\mathrm A})
\propto
P_{\mathrm{net}}(b\mid\mathcal F)P^{\mathrm A}(b),
\end{equation}
where $\mathcal F$ contains the detection instance and previously generated bits. With
$L^{\mathrm A}=\log\left(\frac{P^{\mathrm A}(0)}{P^{\mathrm A}(1)}\right)$ and equal-prior LLR output by the policy network:
\begin{equation}\label{networkodds}
a^{\mathrm{net}}
=
\log\frac{P_{\mathrm{net}}(b=1\mid\mathcal F)}
{P_{\mathrm{net}}(b=0\mid\mathcal F)},
\end{equation}
the posterior LLR and the sampling probability are given as
\begin{align}
\frac{P(b=1\mid\mathcal F,L^{\mathrm A})}
{P(b=0\mid\mathcal F,L^{\mathrm A})}
&=
\exp(a^{\mathrm{net}}-L^{\mathrm A}),
\label{posteriorodds}\\
p^{\mathrm{sample}}
&=
\sigmoid(a^{\mathrm{net}}-L^{\mathrm A}).
\label{priorpolicy}
\end{align}
Thus an a-priori LLR induces a bias into the sampling probabilities of the detector's output. 
At $\bm L^{\mathrm A}=\bm0$, the policy reduces to the previous hard-decision detector.

After the final transition, let $q_{k,j}$ be trajectory $k$'s probability that
bit $j$ is one and let $\bm b_k^{(T)}$ be its sampled vector. We use the exact
normalized residual to combine the $K$ trajectories:
\begin{align}
w_k
&=
\frac{\exp[-f(\bm b_k^{(T)};\xi)/\tau_{\mathrm w}]}
{\sum_{k'=1}^{K}\exp[-f(\bm b_{k'}^{(T)};\xi)/\tau_{\mathrm w}]},
\label{softresidualweight}\\
\bar q_j
&=
\sum_{k=1}^{K}w_kq_{k,j}.
\label{trajectorymarginal}
\end{align}
Because $f$ is normalized by $N_r\sigma^2$, the complex-Gaussian likelihood
satisfies
$p(\bm y\mid\bm b,\bm H)\propto\exp[-N_r f(\bm b;\xi)]$.
Consequently, $\tau_{\mathrm w}=1/N_r$ is the physically matched value in
\eqref{softresidualweight}; any other value is an explicit likelihood-tempering
choice rather than a new estimate of the noise variance. The detector posterior
and extrinsic LLRs are
\begin{equation}\label{posteriorllr}
L_j^{\mathrm P}
=
\log\frac{1-\bar q_j}{\bar q_j},
\quad
L_j^{\mathrm E}
=
L_j^{\mathrm P}-L_j^{\mathrm A}.
\end{equation}
This is a Rao--Blackwellized terminal-mixture estimator: $q_{k,j}$ retains the
final head's bit uncertainty instead of replacing it by the sampled bit, while
the exact likelihood favors candidates that fit the observation.
Only the terminal Bernoulli is Rao--Blackwellized: the mixture weight $w_k$ is
evaluated at the single sampled vector $\bm b_k^{(T)}$ rather than as an
expectation under the terminal head, so \eqref{trajectorymarginal} is a
likelihood-weighted terminal-marginal mixture rather than a fully marginalized
posterior; the two are equal when each trajectory is nearly deterministic.

The extrinsic subtraction in \eqref{posteriorllr} is exact for a single
mean-field trajectory. With $q_{k,j}=\sigmoid(a^{\mathrm{net}}_{k,j}-L^{\mathrm A}_j)$,
\eqref{posteriorodds} gives
$\log((1-q_{k,j})/q_{k,j})=L^{\mathrm A}_j-a^{\mathrm{net}}_{k,j}$, so the
per-trajectory extrinsic $-a^{\mathrm{net}}_{k,j}$ carries only the network
evidence and is free of the prior. After the residual-weighted mixture
\eqref{trajectorymarginal}, however, the logit of a convex combination of
prior-tilted marginals is no longer an affine function of $L^{\mathrm A}_j$, and
the weights $w_k$ themselves depend on $L^{\mathrm A}_j$ through the sampled
vectors. Consequently $L_j^{\mathrm E}=L_j^{\mathrm P}-L_j^{\mathrm A}$ removes
the bit's own prior only approximately, with a residual dependence that shrinks
as the trajectories agree. We therefore measure this residual prior dependence,
rather than assume it is zero, through the prior-sensitivity diagnostic in
Section~\ref{sec:ablation}.

\subsection{IDD Training Scheme}\label{sec:iddtraining}

The main training difficulty introduced by the soft IDD learning scheme is that the detector input prior (i.e., the feedback from LDPC decoder) at each round is not stationary. As detector parameters evolve across the iterations, the distribution of the prior $\bm L^{\mathrm A,(m)}$ also drifts across IDD rounds. Note that the training scheme for the hard-decision receiver assumes equal prior for every transmitted bit and cannot adapt to the practical IDD settings; Moreover, starting from the random initializations (at all IDD rounds) usually renders unstable convergence and overconfident feedback. To address these issues, we propose the following multi-stage training scheme tailored to the soft IDD learning problem.


\subsubsection{Hard-decision initialization}

Note that for the hard-decision receiver in Sec.~\ref{hard_detector_sec}, the same policy $\Pi_{\bm\theta^*_{0}}$ is recursively adopted at all $T$ transitions. This parameter sharing is useful for learning a stable transition rule, but it also constrains the overall expressive power. In contrast, soft-IDD receiver admits a more expressive construction. We unfold the detector across both the $T$ transition steps and the $M$ IDD
rounds \cite{wiesmayr2022duidd}. The resulting parameter family
$\bm\Theta_{\mathrm S}=\{\bm\theta_{\mathrm S}^{(m,t)}\}_{m,t}$ is untied,
so each layer can adapt to its position in the search and to the reliability of the decoder prior received in its round. Crucially, the hard policy provides the common initialization for every untied soft layer, i.e.,
\begin{equation}\label{hardsoftinit}
\begin{aligned}
\left[
\bm P_k^{(m,t)},
\bm Z_k^{(m,t)}
\right]
&=
\Pi_{\bm\theta_{\mathrm S}^{(m,t)}}
\left(
\bm b_k^{(m,t-1)},
\bm Z_k^{(m,t-1)}
\right),
\\
\left.\bm\theta_{\mathrm S}^{(m,t)}\right|_{\mathrm{init}}
&=
\bm\theta_{0}^{\star},
\quad
\substack{m=0,\ldots,M-1,\\t=1,\ldots,T}.
\end{aligned}
\end{equation}
Parameter sharing is released only after initialization: subsequent soft training lets different IDD rounds adapt to increasingly informative decoder
feedback. This gives a direct continuation from the stable, parameter-efficient hard-decision solver to the higher-capacity iterative soft receiver.


\subsubsection{Synthetic-prior training}

The detector is next trained without invoking the LDPC decoder. We draw
synthetic priors from the classical Gaussian LLR model in~\cite{tenbrink2001exit}. For a sampled reliability $\sigma_{\mathrm A}$ and $\varepsilon_j\sim\mathcal N(0,1)$,
\begin{align} \label{syntheticdistribution}
L_j^{\mathrm A}
&=
(1-2b_j^*)\frac{\sigma_{\mathrm A}^{2}}{2}
+\sigma_{\mathrm A}\varepsilon_j, \notag \\
L_j^{\mathrm A}\mid b_j^*
&\sim
\mathcal N
\left(
(1-2b_j^*)\frac{\sigma_{\mathrm A}^{2}}{2},
\sigma_{\mathrm A}^{2}
\right).
\end{align}
Hence the mean has the correct sign for the transmitted bit and its magnitude
increases with reliability. In practice, we sample the target a priori mutual
information and obtain $\sigma_{\mathrm A}=J^{-1}(I_{\mathrm A})$. Covering the required reliability range exposes the detector to uninformative, moderate, and confident priors without decoder-induced drift.

\subsubsection{In-loop training}

The fixed LDPC decoder is inserted in the forward pass, and the exchange in
\eqref{detectorround}-\eqref{priorupdate} is unrolled from
$\bm L^{\mathrm A,(0)}=\bm0$. The resulting priors are exactly those encountered by the deployed receiver, closing the remaining gap from synthetic-prior training.

For training, only the posterior probability from the final transition of
each IDD-round network enters the loss. Let $N_c$ be the number of coded bits
in the interleaved codeword. The round-$m$ posterior BCE is
\begin{align} \label{roundposteriorbce}
\mathcal L_{\mathrm{post}}^{(m)}
= -\frac{1}{N_c} \sum_{j=1}^{N_c} \left[ b_j^*\log\bar q_{m,j} +(1-b_j^*) \log(1-\bar q_{m,j}) \right].
\end{align}
The ultimate soft-IDD training loss is then given by
\begin{align} \label{softloss}
\mathcal L_{\mathrm{soft}} = \sum_{m=0}^{M-1} \omega_m\mathcal L_{\mathrm{post}}^{(m)}, \quad \omega_m\geq0.
\end{align}



The forward pass uses the actual 5G NR LDPC decoder configured according to the
Sionna link-level implementation \cite{hoydis2022sionna}. The module passed to
the next detector is the decoder's \emph{extrinsic} mapping
$\mathcal E_{\mathrm{BP}}(\bm l)=
\mathcal D_{\mathrm{BP}}(\bm l)-\bm l$. During backpropagation, the
straight-through rule is applied to this composite mapping:
\begin{equation}\label{decoderste}
\frac{\partial
\mathcal E_{\mathrm{BP}}(\bm l)}
{\partial\bm l}
\approx
\bm I.
\end{equation}
The true decoder posterior and the subtraction are therefore both retained in
the forward pass, whereas the backward pass copies the gradient through the
whole extrinsic module.


\section{Performance Evaluation}\label{sec:perfecteval}

\subsection{Simulation Settings}\label{sec:settings}

We consider a downlink MIMO system with perfect CSI available at the receiver. Each channel sample follows the Rayleigh distribution, with $[\bm{H}]_{ij} \overset{\text{i.i.d.}}{\sim} \CN(0,1/N_t)$, and Gray-coded 64-QAM or 256-QAM symbols are normalized to unit average energy. Unless otherwise stated, the L2O framework uses $T=8$ transitions, embedding dimension $d=256$, eight attention
heads, feed-forward dimension $d_{\mathrm{ff}}=256$, and $K=16$ trajectories.
One trajectory is initialized by the LMMSE estimate and the remaining
trajectories by its randomized perturbations. Residual-only post-processing is
disabled so that the reported gain is attributable to the learned transition
policy.

The hard detector is trained at 20~dB for $10^4$ AdamW updates with learning
rate $10^{-4}$. The residual-to-BER curriculum uses $s_0=10^3$ and
$s_1=5\times10^3$, and the batch sizes are 64 and 16 for 64-QAM and 256-QAM,
respectively. For soft reception, every unrolled detector layer is initialized
from the same hard checkpoint according to \eqref{hardsoftinit} and then follows the synthetic-prior and
decoder-in-the-loop stages in Section~\ref{sec:iddtraining}. Soft training uses $7\times10^3$
updates, learning rate $3\times10^{-5}$, and batch size 16, with $K=32$ for
64-QAM and $K=16$ for 256-QAM.

The coded link uses the 5G NR LDPC implementation in Sionna
\cite{hoydis2022sionna}, code rate $1/2$, ten belief-propagation iterations,
and three IDD rounds. Each codeword packs $P=8$ independently detected MIMO
vectors, giving rate-matched codeword lengths of 384 and 512 bits for 64-QAM
and 256-QAM, respectively. Every reported SNR point is averaged over at least
five random seeds until either 200 frame errors or $10^7$ information bits are
observed.

For the hard detector, we compare LMMSE, QR-domain K-best, OAMP-Net2, RE-MIMO,
and SGT \cite{he2020model,pratik2021remimo,hong2025sgt}. For soft IDD, the
comparisons are soft LMMSE, MMSE-PIC, K-best list-MAP, SGT, and DUIDD
\cite{wiesmayr2022duidd}. All methods use the same channel realizations,
modulation, code, decoder iterations, and SNR definition. Uncoded BER is the
primary hard-decision metric. Post-decoding BLER is the primary soft-receiver
metric, with information-bit BER, generalized mutual information (GMI), and
Brier score used to diagnose the quality of the detector LLRs. Parameter count
and batch-one latency are measured under identical hardware, software,
precision, and synchronization settings; detector and decoder latency are
reported separately.

\subsection{Ablation Studies}\label{sec:ablation}

We first isolate the residual-to-BER curriculum at one representative 64-QAM
SNR near the BER waterfall. The proposed smooth schedule is compared with
residual-only training, BCE-only training, and an abrupt residual-to-BCE switch.
For each variant, we report final BER, normalized residual, and the standard
deviation over five seeds. For the full model, BER, residual, and normalized
Hamming distance between consecutive vectors are also plotted against the
transition index. This compact experiment tests both roles of the curriculum:
the residual phase should establish a stable search direction, whereas the BCE
phase should align the terminal policy with the transmitted bits. Across
transitions, a useful learned policy should reduce BER and residual without
collapsing immediately to identical trajectories.


The soft-receiver ablation is performed at one SNR near the coded waterfall and
is summarized in a single table. Starting from the complete three-stage
training scheme, we separately (i) remove the hard-checkpoint initialization,
(ii) retain hard-style parameter tying across all soft-detector layers after the common initialization,
(iii) remove prior tilting from candidate generation, (iv) omit the
synthetic-prior stage, (v) omit decoder-in-the-loop fine-tuning, and (vi) stop
the gradient at the LDPC decoder instead of using \eqref{decoderste}. We report
BLER after each IDD round together with final-round GMI and Brier score. To
check that prior subtraction in \eqref{posteriorllr} does not leave excessive
self-information, we additionally vary one bit's input prior while holding the
channel observation and all other priors fixed. We then report the sensitivity
of that bit's extrinsic LLR and the rate at which its sign changes. This targeted
test diagnoses the approximation in \eqref{posteriorllr} without requiring a
full EXIT characterization.


\subsection{Overall Performance}\label{sec:overall}

The first main experiment reports uncoded BER versus SNR for 64-QAM and
256-QAM in two panels. This comparison determines whether the learned
complete-vector transitions improve upon linear detection and existing neural
detectors, and whether they approach the K-best performance obtained with an
explicit tree search.


The second main experiment reports post-LDPC BLER versus SNR for both
modulation orders. Curves after one, two, and three IDD rounds are shown for the
proposed receiver; the baselines are shown after the same number of decoder
iterations and, when applicable, the same number of detector--decoder
exchanges. Final-round information-bit BER and detector GMI are reported in a
small companion table. The round-by-round curves directly test whether
Bayesian prior-tilted sampling converts decoder feedback into increasingly
informative candidate sets rather than merely recalibrating a fixed hard
decision.




\end{document}